A NOTE ON BLACK HOLE TEMPERATURE AND ENTROPY

P. R. SILVA

Departamento de Física – Instituto de Ciências Exatas – Universidade Federal de Minas Gerais – C. P. 702 – 30123-970 –Belo Horizonte – Minas Gerais – BRAZIL
e-mail: prsilva@fisica.ufmg.br

ABSTRACT- We propose intuitive derivations of the Hawking temperature and the Bekenstein-Hawking entropy of a Schwarzschild black hole.

In a very interesting paper entitled : " The quantum black hole ", Ram and collaborators [1, 2] state that : " the quantum nature of a black hole is revealed using the simplest terms that one learns in undergraduate and begining graduate courses." Ram and co-workers close their paper answering the question: what is the quantum nature of a black hole? And they conclude: " A black hole is a Bose-Einstein ensemble of quanta of mass equal to twice the Planck mass, confined in a sphere of radius twice the black hole mass."

The concept of black hole entropy was introduced by Bekenstein [3] which stated that it is proportional to the black hole area A. Later on, Bardeen, Carter and Hawking [4] and Hawking [5], performed calculations using a semi-classical approximation, putting Bekenstein conjecture on a firm basis. They established that the black hole temperature is proportional to its surface gravity.

On the other hand, A. Zee [6] in his book: "Quantum Field Theory in a Nutshell", presents a nice intuitive derivation of the Hawking temperature. Zee starts from the Schwarzschild solution of Einstein equations for vacuum, namely

$$ds^2 = (1 - 2GM/r)\,dt^2 - (1 - 2GM/r)^{-1}\,dr^2 - r^2 d\theta^2 - r^2 \sin^2\theta\, d\Phi^2. \quad (1)$$

By setting $t \to -i\tau$, he performs Wick rotation on the metric obtaining

$$ds^2 = -[(1 - 2GM/r)\,d\tau^2 + (1 - 2GM/r)^{-1}\,dr^2 + r^2 d\theta^2 + r^2 \sin^2\theta\, d\Phi^2]. \quad (2)$$

The interest is to examine the solution just outside the horizon, namely $r > 2GM$. Now Zee proposes the change of coordinates

$$R\,d\alpha = (1 - 2GM/r)^{1/2}\,d\tau, \quad (3)$$

and

$$dR = (1 - 2GM/r)^{-1/2}\,dr. \quad (4)$$

Integrating both equations, where he keeps in the solutions leading order terms in $(r - 2GM)$, and by taking $\alpha$ going from 0 to $2\pi$, $\tau$ from 0 to $\beta$ and $r$ from $2GM$ to $r$, he gets:

$$R\,2\pi \approx (2GM)^{-1/2}(r - 2GM)^{1/2}\,\beta, \quad (5)$$

and

$$R \approx (2GM)^{1/2} \, 2 \, (r - 2GM)^{1/2}. \quad (6)$$

Dividing (5) by (6), he obtains after considering that $\beta = 1/T$, the expression for the Hawking temperature, namely

$$T = 1/(8\pi GM) = \hbar c^3/(8\pi GM), \quad (7)$$

where $\hbar$ and $c$ were restored in the right side of (7).

It is the purpose of this note to derive the black hole temperature and entropy in an alternative way to that followed by Zee [6] and also Ram [1,2]. Let us consider at the surface horizon of a black hole a circle which perimeter is equal to $2\pi R_S$, where $R_S$ is the Schwarzschild radius of a body of mass M, located at its center (see figure 1). A fluctuating field which can radiate, will be considered as a transverse field $g_t$ and we suppose that it has equal probability to assume positive or negative values. A test particle of mass m coupled to this field, integrated over a small interval of the circle's arc $\Delta S$ will give on average null contribution for the energy, due to the fluctuating character of this field. Thus we need to take into account the second moment of this "elastic" energy, and we write

Figure 1

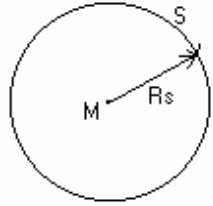

$$\Delta U = \frac{m^2 \, g_t^2 \, (\Delta S/2\pi)^2}{2mc^2}. \quad (8)$$

In (8), we have "normalized" $\Delta U$, dividing the numerator by $2mc^2$ by taking into account the possibility of producing a particle pair from the vacuum. The field $g_t$ is assumed to be responsible by the black hole evaporation. We want to compare this field with the radial field $g_r$, which maintains the test particle tied to the black hole. In (8), we also rescaled $\Delta S$ dividing it by $2\pi$. In this way we assure that a light front will take the same time of travel to go from the center of the sphere to its surface horizon as to cover its rescaled perimeter.

We can also write

$$\Delta U = 1/2 \, k \, (\Delta S)^2, \quad (9)$$

where

$$k = m g_t^2 / [(2\pi)^2 c^2], \quad (10)$$

is a spring constant of a harmonic oscillator. The angular frequency of this oscillator is given by

$$\omega = (k/m)^{1/2} = g_t / (2\pi c), \quad (11)$$

and yields to

$$g_t = 2\pi\omega c. \quad (12)$$

On the other hand, the radial field at the surface horizon is given by

$$g_r(r=R_S) = GM/(R_S^2) = c^4/(4GM), \quad (13)$$

where $R_S = 2GM/c^2$ is the Schwarzschild radius. We propose that at the surface horizon the vacuum fluctuations are so great that the strength of the radiative field equals to the radial binding field. So, by imposing the equality between $g_t$ and $g_r$ fields given by (12) and (13) respectively, we get

$$\omega = c^3/(8\pi GM), \quad (14)$$

and

$$k_B T = \hbar\omega = \hbar c^3/(8\pi GM), \quad (15)$$

which is just the Hawking temperature found in (7), if we take the Boltzmann constant equal to one.

In order to obtain the Bekenstein-Hawking entropy let us take the free energy of a black hole as the thermodynamic relation

$$F = E - TS, \quad (16)$$

where E and F are respectively the internal and free energies and S is the entropy. Let us take the extremum of F, by considering an isothermal process. We have

$$\Delta F = \Delta E - T\Delta S = 0. \quad (17)$$

To evaluate $\Delta E$, we consider a black hole interacting with itself and think about a body of reduced mass equal to M/2. With these ideas in mind we can write

$$\Delta E = \tfrac{1}{2} Mc^2. \quad (18)$$

Substituting T, given by (15) and $\Delta E$, given by (18), in (17) and after solving for $\Delta S$, we obtain

$$\Delta S = (k_B 4\pi GM^2)/(\hbar c), \quad (19)$$

or

$$\Delta S = (k_B/4)(4\pi R_S^2)/(L_{Pl}^2), \quad (20)$$

where $L_{Pl}^2 = \hbar G/(c^3)$ is the square of the Planck radius. Taking

$$S = S_0 + \Delta S, \quad (21)$$

we can recover the well known result of Bekenstein-Hawking entropy, by setting $S_0 = 0$.

B. Ram and co-workers [1] took the the quantum harmonic oscillator to model the quantum black hole and then used the Bose statistics as a means to evaluate its entropy.

In the present work we are also taking in account a quantum harmonic oscillator model, but it differs from that considered in the work of B. Ram and collaborators. To be specific, in our work the separation in energy levels $\hbar\omega$ is given by equation (15), while the separation in the Ram et al model is twice the Planck mass. In the following we propose that the "self- interaction" of the black hole will correspond to the onset of $\eta$ elementary excitations, namely:

$$\eta\hbar\omega = \Delta E = \tfrac{1}{2} Mc^2. \quad (22)$$

Putting $\hbar\omega$ given by (15) into (22) and solving for $\eta$, we obtain

$$\eta = 4\pi R_S^2/(4L_{Pl}^2). \quad (23)$$

In (23), we interpret $\eta$ as the number of cells of length equal to twice the Planck length contained in the area of the surface event horizon of the black hole.

Now let us suppose the existence of a particle that we will call planckion. We propose that there is a one to one correspondence between the planckions and the the elementary excitations of energy $\hbar\omega$. The planckion has mass equal to the Planck mass and quantum length equal to the Planck length. Each planckion lives on its proper cell and interacting with itself it reduces to half its mass-energy, therefore enlarging its length to twice the Planck length. The planckion can not leave its cell, due to its finite lifetime imposed by the uncertainty principle. Inside its proper cell it has at its disposal four (4) microstates. This situation differs from that which occurs with a molecule of an ideal gas, which could in principle visit all the cells of the volume occupied by the gas, if we wait a sufficient long time.

By considering the ensemble of $\eta$ cells, each with 4 microscopic states, we can write the total number of microstates w as

$$w = 4^\eta. \quad (24)$$

Using Boltzmann formula, we have for the entropy of the black hole:

$$S = k_B \ln w = k_B \eta \ln 4 = \ln 4\, k_B\, 4\pi R_S^2/(4L_{Pl}^2). \quad (25)$$

This approximated estimate of the black hole entropy differs from its accurate result by the presence in (25) of a factor $\ln 4$, instead of $\ln e = 1$.

To pursue further we notice that relation (22) can be writen in the form

$$\eta = 4\pi M^2/(M_{Pl}^2). \quad (26)$$

On the other hand B. Ram et al [1] assume that the black hole mass contains 2N quanta of elementary Planck mass, namely

$$M = 2N\, M_{Pl}. \quad (27)$$

Comparing (26) and (27) we get:

$$N = 1/4\, (\eta/\pi)^{1/2}. \quad (28)$$

Therefore, the number of cells which cover the horizon area of a black hole is proportional to the square of the number of the mass quanta established by B. Ram and co-workers [1].

To close this note we judge worth to refer to a more extensive list of bibliographic references. This can be found in a paper by T. Damour [7] entitled: "The entropy of black holes: a primer".

ACKNOWLEDGEMENTS- We are grateful to Domingos Sávio de Lima Soares and Carlos Heitor d'Ávila Fonseca for helpful discussions and critical readings of the manuscript. We also thank to Antônio Sérgio Teixeira Pires for helpful discussions.